\documentclass[twocolumn,showpacs,preprintnumbers,amsmath,amssymb,nofootinbib,floatfix]{revtex4}

\usepackage{graphicx}% Include figure files
\usepackage{dcolumn}% Align table columns on decimal point
\usepackage{bm}% bold math

\newcommand{\SUSY}{{ \it SUSY}}

\begin{document}

\preprint{
\vbox{
      \hbox{August 2007} }}

\title{Decoupling and Destabilizing in Spontaneously Broken Supersymmetry}
\author{Jonathan A.~Bagger and Adam F.~Falk}
\affiliation{Department of Physics and Astronomy, The Johns Hopkins
University\\ 3400 North Charles Street, Baltimore, Maryland
21218}

\thispagestyle{empty}
\begin{abstract}%
The supersymmetric analog of the Goldberger-Treiman relation plays a critical role in the low energy effective theory of models in which supersymmetry is spontaneously broken in a hidden sector.  The interactions that connect the hidden and visible sectors break a global symmetry, which implies that the low energy theory must be constructed consistently in inverse powers of the messenger scale.  The Goldberger-Treiman relation determines the couplings of the Goldstino to the visible sector fields.  These couplings are fixed by the soft supersymmetry breaking terms within a power counting scheme that is stable under radiative corrections.  We describe the power counting of the low energy effective theory, first for a toy model of extended technicolor and then for the supersymmetric standard model.  One implication of this work for supersymmetry phenomenology is the observation that Goldstino loops can destabilize the weak scale if the low energy theory is not constructed consistently.  Another is that Goldstino loops induce all visible sector operators not forbidden by symmetries.  The magnitudes of these operators are determined by the consistent power counting of the low energy effective theory.
\end{abstract}


\maketitle

\newpage

\section{Introduction}

The primary purpose of this paper is to make an elementary but essential observation about the qualitative features of models with spontaneously broken supersymmetry in a hidden sector.  We will argue that the visible sector soft supersymmetry breaking terms must obey a self-consistent power counting if the weak scale is not to be destabilized.  This power counting is the one that follows from the fact that an explicit symmetry-breaking parameter controls all communication between visible  and hidden sector fields.  Because of the Goldberger-Treiman relation~\cite{Goldberger:1958tr}, the same power counting simultaneously governs both the soft symmetry breaking terms and the Goldstino interactions.  We will show that Goldstino loops enforce the correct power counting by introducing destabilizing divergences in an incorrectly constructed low energy  theory.  We will also see that Goldstino loops induce all operators in the low energy effective theory not forbidden by symmetries, with coefficients determined by the consistent power counting.

An immediate consequence of our work concerns the minimal supersymmetric standard model (MSSM).  In Refs.~\cite{Hall:1990ac,Weinberg:2000cr} it was noted that nonholomorphic scalar trilinears can be added to the soft supersymmetry breaking lagrangian with weak scale coefficients,
\begin{equation}
m_W  q \bar u h^*_d + m_W  q \bar d h^*_u\,,
\end{equation}
where $h_u$ and $h_d$ are Higgs fields, and $q$, $\bar d$ and $\bar u$ are squarks.  According to the usual power counting, these terms are soft, so they should not destabilize the hierarchy.  In this paper we show that this argument is correct, but incomplete.  When Goldstinos are included, the coefficients of the nonholomorphic trilinears must be of order $m_W^2/M$ (where $M$ is the messenger scale), or Goldstino loops will destabilize the hierarchy. 

A secondary purpose of this paper is to explore more generally the low energy effective theory of the visible sector in models with hidden sector symmetry breaking.  We assume that the symmetry breaking at a high scale $f$ is communicated to the visible sector through nonrenormalizable interactions suppressed by an even higher scale $M$.  We will argue that because the global symmetry of the (visible + hidden) theory is enhanced as $M\to\infty$, one can consistently count powers of $1/M$ in the visible sector, even below the scale $f$ at which dynamics in the hidden sector may become strongly coupled.  As a consequence, in the low energy  theory of the visible sector the ultraviolet cutoff may be taken as high as $M$, while the coefficients of nonrenormalizable operators, suppressed by powers of $1/M$, may depend on nonperturbative hidden sector dynamics or logarithms of $f/M$.  These terms look, from a visible sector point of view, like matching terms generated at $M$,  but they actually depend on physics at the lower (but still high) scale $f\ll M$.  

Because these more general features are not particular to supersymmetry, we will first discuss them in the somewhat simpler context of a toy model of extended technicolor.  We will then apply the lessons learned to supersymmetry, and conclude with an analysis of the MSSM.

\section{Effective Field Theory of an Extended Technicolor Model}

This paper focuses on power counting in models with hidden sector supersymmetry breaking.  Nevertheless, almost all of the relevant physics can be understood by studying a simple toy model of extended technicolor (ETC).  We will develop this example in considerable detail and map the conclusions of our discussion onto supersymmetry at the end.  

The ETC model has a ``visible sector'' consisting of quark fields $Q_L$ and $Q_R$, transforming under a global $SU(2)_{LQ}\times SU(2)_{RQ}$:
\begin{equation}
  Q_L\to L_Q Q_L\,,\qquad Q_R\to R_Q Q_R\,.
\end{equation}
It also has a ``hidden sector'' with techniquarks $T_L$ and $T_R$, transforming under a separate global chiral symmetry $SU(2)_{LT}\times SU(2)_{RT}$,
\begin{equation}
  \overline T_L\to L_T \overline T_L\,,\qquad \overline T_R\to R_T \overline T_R\,.
\end{equation}
The two sectors are coupled by a nonrenormalizable contact term that arises from extended technicolor at a high scale $M$, 
\begin{equation}\label{ETCint}
  {\alpha\over M^2}\,\overline Q_L \overline T_L\,T_R Q_R\,.
\end{equation}

The parameter $\alpha/M^2$ may be viewed as a ``spurion,'' in the sense that it is a constant that carries $SU(2)^4$ quantum numbers.  The ETC interaction~(\ref{ETCint}) breaks $SU(2)_{LQ}\times SU(2)_{LT}\to SU(2)_L$ and $SU(2)_{RQ}\times SU(2)_{RT}\to SU(2)_R$, leaving the $SU(2)_{L}\times SU(2)_{R}$ subgroup intact.  We put scare quotes around ``spurion'' because we do not necessarily view $\alpha$ as representing the vacuum expectation value of a field.  This spurion (acknowledging this possible abuse of terminology, we now drop the quotes) carries a power counting dimension of $1/M^2$; most importantly, it alone controls the $SU(2)_{LQ}\times SU(2)_{RQ}\times SU(2)_{LT}\times SU(2)_{RT}\to SU(2)_L\times SU(2)_R$ symmetry breaking in the low energy effective theory.

At some scale $f$, the technicolor interaction becomes strong and the techniquarks condense, $\langle\overline T_L T_R\rangle= f^3$.  (In this paper we are interested only in the general question of power counting the mass scales present in the theory, so we drop all factors of $4 \pi$.  Of course, the correct counting of such loop factors may be phenomenologically important in a particular theory.)  This condensate breaks the $SU(2)_{LT}\times SU(2)_{RT}$ symmetry to its diagonal subgroup, and produces three Goldstone bosons in the effective theory below $f$.  The product $\overline T_L T_R$ is replaced by a composite field containing the Goldstone multiplet $\Pi=\pi^a T^a$,
\begin{equation}
  \overline T_L T_R\to  f^2\Sigma
  =f^3 e^{i\Pi/f}.
\end{equation}
This is in accord with the general construction of Callan, Coleman, Wess and Zumino~\cite{Coleman:1969sm,Callan:1969sn}, which follows simply from the structure of the spontaneously broken symmetry.  The field $\Sigma$ transforms as $\Sigma\to L_T\Sigma R_T^\dagger$ under $SU(2)_{LT}\times SU(2)_{RT}$.  In terms of $\Sigma$, the ETC interaction becomes
\begin{equation}\label{etcgold}
  {\alpha f^2\over M^2}\,\overline Q_L \Sigma Q_R\,.
\end{equation}

The interactions of the quarks with the Goldstones are found by expanding $\Sigma$.  The leading term gives rise to the quark mass $m=\alpha f^3/M^2$, which breaks the quark sector global symmetry $SU(2)_{LQ}\times SU(2)_{RQ}$ to the diagonal $SU(2)$ subgroup.  The breakdown of chiral symmetry in the quark sector is the result of chiral symmetry breaking in the techniquark sector, communicated by a nonrenormalizable interaction.  Note that breaking the full symmetry $SU(2)_{LQ}\times SU(2)_{RQ}\times SU(2)_{LT}\times SU(2)_{RT}\to SU(2)_L\times SU(2)_R\to SU(2)$ requires {\it both\/} symmetry breaking parameters: the condensate $\langle\overline T_LT_R\rangle$ and the spurion $\alpha/M^2$.  

Expanding the interaction (\ref{etcgold}) to linear order in $\Pi$, 
\begin{equation}
  m\overline Q_LQ_R+i{m\over f}\,\overline Q_L\Pi Q_R+\dots\,,
\end{equation}
we see that the chiral symmetry breaking mass term is related to the Goldstone boson coupling via $m=g_{\pi QQ}f$.  This is the famous Goldberger-Treiman relation.  It follows solely from the fact that the chiral symmetry is broken spontaneously.  Note that the Goldstone-quark coupling is suppressed by $m/f\sim f^2/M^2$.

It is instructive to consider the power counting in this theory from a top-down, ``Wilsonian'' point of view.  From this perspective, one starts with a theory defined at short distances and then progressively integrates out modes of longer and longer wavelength to arrive at an effective theory at lower energies.   At each energy scale, the relevant symmetries are those that are unbroken at that scale.  The natural sizes of the operator coefficients are just the magnitudes that are generated by this procedure.  Delicate cancellations between coefficients present at short distances and corrections generated by Wilsonian running are considered to be unnatural.

Above the ETC scale $M$, we assume that we have a theory in which the techniquark and quark fields interact with each other freely.  (The theory could be nonrenormalizable, but any nonrenormalizable operators at that scale give even smaller contributions at low energies than the ones we are interested in.)  When we integrate out modes at and above the scale $M$, including the ETC sector of the theory, we generate a tower of operators suppressed by powers of $1/M$.  Below $M$, these nonrenormalizable operators provide the only interactions between the quark and techniquark sectors.

We now integrate out momenta between $M$ and the lower scale $f$.  All modes have energies below $M$, so no powers of $M$ are induced in the numerators of the operator coefficients.  Thus the organization of operators according to power counting in $1/M$ is preserved.  Of course, this is the standard behavior of an effective field theory.

The final step is to run the theory from $f$ to $m\sim f^3/M^2$.  In this momentum regime, the degrees of freedom of the techniquark sector are those of a theory with spontaneously broken $SU(2)_{LT}\times SU(2)_{RT}$.  They contain the massless nonlinearly self-coupled Goldstone multiplet $\Sigma$.  There may be other light particles in the techniquark sector, but whether they are strongly or weakly coupled below $f$, they do not change the Goldstone boson power counting in the quark sector.  

The Goldstone multiplet contains the chiral symmetry breaking term $f$.  This chiral symmetry breaking is communicated to the quark sector through $1/M$ suppressed operators that break $SU(2)_{LQ}\times SU(2)_{RQ}\times SU(2)_{LT}\times SU(2)_{RT}\to SU(2)_L\times SU(2)_R$.  Because of these operators, the quark sector ``sees'' a techniquark sector in which chiral symmetry is realized nonlinearly.  The weakly coupled spurion insertions transmit the chiral symmetry breaking to the quark sector.

Running from $f$ down to $m$, it is impossible to generate effects that produce positive powers of the scale $M$.  The effective lagrangian for the field $\Sigma$ is not calculable perturbatively, but at the scale $m$ it remains a nonlinear function of $\Pi$ and $\partial$.  (The back reaction from the quark sector could induce corrections suppressed by powers of $1/M$, but these are comparatively tiny and unimportant.)  Meanwhile, the quark sector modes that are integrated out are no different than they were between $f$ and $M$.  Running from $f$ to $m$ changes the coefficients of the nonrenormalizable operators, but it does not change the power counting in $M$ and $f$.

From the Wilsonian point of view, the effective lagrangian at the scale $m$ receives contributions from a variety of short distance modes that have been integrated out of the theory.  These include quark modes with momenta between $m$ and $M$, Goldstone modes between $m$ and $f$, and techniquark modes between $f$ and $M$.  Of course, we would not actually calculate with a Wilsonian effective action containing a strict ultraviolet cutoff.  Instead, we would use a momentum-independent scheme such as dimensional regularization and minimal subtraction.  For that reason, our effective theory would have loop divergences that must be subtracted.  The value of the Wilsonian perspective is the insight it gives into the physical meaning of these divergences.

To determine the correct power counting, it is useful to recall the various types of ultraviolet behavior that can appear when the operators in the effective theory, some of which are nonrenormalizable, are put into loops.  Ultraviolet convergent diagrams are dominated by momenta at or below the scale of the effective theory, and are calculable.  Logarithmically divergent diagrams  receive contributions from momenta at all scales and are interpreted as controlling the scale dependence of the effective operators.  Finally, power divergent diagrams are dominated by momenta at the cutoff scale; they contain no information about the low energy  dynamics.  These divergences are cancelled by matching corrections at the high scale.  The corresponding counterterms necessarily obey a consistent power counting in inverse powers of the cutoff. 

What is the power counting in the case of our ETC toy model?  The essential question is the scale at which power divergences should be cut off.   The answer depends on the fields that appear in the loops.  Goldstone loops are cut off at $f$ because above this scale, the Goldstones must be replaced by the techniquarks as the appropriate degrees of freedom.  On the other hand, loops that contain only quarks run all the way up to the messenger scale $M$.  The power divergences determine the power counting of the counterterms, and we have argued that the coefficients in the effective theory receive contributions from quark sector modes with momenta in the entire range from $m$ to $M$.

These are the conditions for evaluating loops that are consistent with the rules for power counting given earlier.  Such consistency conditions are analogous to those given by Georgi and Manohar~\cite{Manohar:1983md} for the effective theory of chiral quarks and pions.

Note that under the chiral $SU(2)_L \times SU(2)_R$, the quark fields transform linearly and the Goldstone fields nonlinearly.  In this formulation, the decoupling of the Goldstone bosons as $M\to\infty$ is manifest, but the Goldstone bosons are not explicitly derivatively coupled.  (The absence of a momentum-independent contact interaction can be shown using the equations of motion.) However, one can perform a nonlinear field redefinition, leaving the physical content of the effective theory invariant, that makes the derivative coupling manifest~\cite{Coleman:1969sm,Callan:1969sn}.  This is done by writing $\Sigma=\xi^2$, and redefining
\begin{equation}
  \widehat Q_L=\xi^\dagger Q_L\,,\qquad \widehat Q_R=\xi Q_R\,.
\end{equation}
After this transformation the symmetry breaking mass term is simply $m\overline {\widehat Q}_L \widehat Q_R$, but the kinetic terms now take the form
\begin{eqnarray}
  \overline Q_L\,\gamma^m\partial_m Q_L&=&
  \overline{\widehat Q}_L\,\gamma^m\left(\partial_m
  +\xi^\dagger\partial_m\xi\right)\widehat Q_L\,,\cr
  \overline Q_R\,\gamma^m\partial_m Q_R&=&
  \overline{\widehat Q}_R\,\gamma^m\left(\partial_m
  +\xi\partial_m\xi^\dagger\right)\widehat Q_R\,.
\end{eqnarray}
For our purposes, this formulation obscures the essential points.  The Goldberger-Treiman relation is hidden, and is only revealed when the equations of motion are applied to matrix elements.  The decoupling of the Goldstone bosons as $M\to\infty$ is not manifest.  It is also difficult to identify the appropriate cutoffs for the different types of loop diagrams.

\section{Power Counting in Models with a Hidden Sector }

The ETC model described above is directly relevant to our discussion of hidden sector supersymmetric models.  In each case, the symmetry is broken in a ``hidden'' sector and then communicated to a ``visible'' sector via interactions suppressed by powers of $1/M$.  Recalling that the quark and techniquark parts of the ETC model play the roles, respectively, of the visible and hidden sectors, we note that:

$\bullet$ The global symmetry of the renormalizable interactions is enhanced to separate global symmetries in the visible and hidden sectors.

$\bullet$ Nonrenormalizable operators, suppressed by the scale $M$, connect the visible and hidden sectors and break the enhanced global symmetry.  Below the scale $M$, the coefficients of these operators may be viewed as spurion fields with a definite power counting.  The decoupling of the two sectors as $M\to\infty$ is associated with the restoration of the enhanced symmetry.

$\bullet$ Below the scale $f$, there is a Goldberger-Treiman relation between the symmetry breaking terms in the visible sector and the coupling of visible sector fields to the Goldstone bosons.  Consistent with the Goldberger-Treiman relation, the Goldstone bosons decouple from the visible sector as $M\to\infty$.

$\bullet$ The effective theory below $f$ is constructed so that it is consistent and stable under renormalization; the suppression of the $\Pi$ field by $f$ is consistent with cutting off Goldstone loop momenta at the scale $f$.  Conversely an effective theory that is {\it not\/} constructed consistently will not be stable under radiative corrections.  The breaking of the entire global symmetry must be power counted correctly term by term.

$\bullet$  In loops with Goldstone bosons, momenta are cut off at the chiral symmetry breaking scale $f$.  In quark loops, momenta are cut off at the messenger scale $M$.

$\bullet$ One could integrate the techniquark sector out of the theory entirely, producing a theory of only quark fields. (This theory would be nonlocal because the Goldstones are massless.)  The ultraviolet cutoff for such a theory would be $M$.  What is unusual is that the coefficients of nonrenormalizable operators, suppressed by powers of $1/M$, typically depend on hidden sector dynamics at the lower scale $f$, or on logarithms of $f/M$.\footnote{An example, in a supersymmetric context, of how such hidden sector logarithms can appear in visible sector operators is presented in Ref.~\cite{Cohen:2006qc}.}  The point is that because the global symmetry of the entire theory is enhanced as $M\to\infty$, one can consistently count powers of $1/M^2$ in the visible sector, even below the scale $f$ at which dynamics in the hidden sector may become strongly coupled.

These features all have analogs in the case of supersymmetry breaking.  First, consider the global symmetries of a supersymmetric theory with visible and hidden sectors coupled only by interactions suppressed by a messenger scale $M$.  As $M\to\infty$, the two sectors decouple from each other, and the global supersymmetry is enhanced to $\SUSY_V\times \SUSY_H$.  (To be clear, this is $\it not$ $N=2$ supersymmetry, because the generators of $\SUSY_V$ and $\SUSY_H$ close into Hamiltonians acting, respectively, only on the visible  and hidden sector fields.  But these two Hamiltonians commute in the decoupling limit; in this limit, and neglecting gravity, the hidden and visible sectors can actually be thought of as functions of independent sets of spacetime coordinates!)  

The terms suppressed by $M$ break the global symmetry to ordinary supersymmetry, $\SUSY_V\times \SUSY_H\to \SUSY$; they should be thought of as containing a spurion field that transforms under $\SUSY_V\times \SUSY_H$ and acquires a vacuum expectation value that leaves only ordinary supersymmetry unbroken.  If supersymmetry is broken in the hidden sector at some scale $\sqrt F$, it is really $\SUSY_H$ that is broken.  It only becomes $\SUSY_V$ breaking in the presence of the spurion.

Hence supersymmetry breaking in the visible sector is always accompanied by a spurion that contains explicit factors of $1/M$.  Since the spurion terms vanish as $M\to\infty$ for any value of $F$, supersymmetry breaking effects can always be decoupled.  To preserve this decoupling, any consistent power counting of supersymmetry breaking at low energies must contain a consistent counting in powers of $1/M$.

\section{Power Counting in Models with Broken Supersymmetry}

We now turn to the construction of the effective lagrangian for spontaneously broken supersymmetry.  We assume that supersymmetry is broken in a hidden sector by some chiral superfield that acquires a nonzero $F$ term $F\theta\theta$.  The precise nature of this field, whether it is elementary or composite, is irrelevant to the low energy  physics of the visible sector.  We do know one other thing about it, however: spontaneous breaking of global supersymmetry implies the existence of a massless Goldstino, which by the construction of Wess and Samuel~\cite{Samuel:1982uh} may be assembled into an analog of $\Sigma$ from the ETC example above.  This field is $S=F\Theta\Theta$, where
\begin{equation}\label{Theta}
  \Theta_\alpha=\theta_\alpha+{1\over F}\,G_\alpha+
  {2 i \over F^2}\,\theta\sigma^m\overline G\,\partial_m G_\alpha
  +\dots
\end{equation}
is a chiral superfield that contains the Goldstone fermion $G$.

Given a visible sector chiral superfield $\Phi$, one may write an interaction term, analogous to (\ref{ETCint}), that connects the visible and hidden sectors,
\begin{equation}
{\alpha\over M^2}  \int {\rm d}^4\theta\,
  S S^\dagger \Phi^\dagger\Phi
  =m^2\int {\rm d}^4\theta\,
  \Theta\Theta\overline\Theta\overline\Theta\,\Phi^\dagger\Phi\,.
\end{equation}
Here $m^2=\alpha F^2/M^2$ is a soft supersymmetry breaking mass.  The coefficient $\alpha/M^2$ of this operator is the spurion that breaks $\SUSY_V\times \SUSY_H\to \SUSY$.  It is clear that as $M\to\infty$ the enhanced symmetry is restored.  Writing $\Phi=A+\theta\chi+\dots$, and expanding $\Theta$ to linear order in $G$, we see that this term also gives rise to a Goldberger-Treiman relation for spontaneously broken supersymmetry.  Here, the soft mass term $m^2A^*A$ is related to the Yukawa coupling $(m^2/F)A^*\chi G$.

The Goldberger-Treiman relation follows from the fact that supersymmetry is spontaneously broken.  There is an analogous relation for every operator that communicates supersymmetry breaking to the visible sector~\cite{Girardello:1981wz}.  Such terms include soft masses,
\begin{equation}\label{softmass}
  {1\over M^2}\int {\rm d}^4\theta\,
  S S^\dagger \,\Phi^\dagger\Phi
  \Rightarrow m_0^2\,A^*A+
  {m_0^2\over F}\, A^*\chi G+\dots\,,
\end{equation}
\begin{equation}\label{softBmass}
 {1\over M^2}\int {\rm d}^4\theta\,
  S S^\dagger \,\Phi^2
  \Rightarrow B_0^2 \,A^2+
  {B_0^2 \over F}\, A \chi G+\dots\,,
\end{equation}
gaugino masses,
\begin{equation}
  {1\over M}\int {\rm d}^2\theta\,
  S W^\alpha W_\alpha\Rightarrow 
   m_{1/2}\,\lambda\lambda+
  {m_{1/2}\over F}\,\lambda\sigma^{mn} GF_{mn}+\dots\,,
\end{equation}
and scalar self-interactions,
\begin{equation}
  {1\over M}\int {\rm d}^2\theta\,
  S\,\Phi^3\Rightarrow  A_0\,A^3+
  {A_0\over F}\,A^2\,\chi G+\dots\,,
\end{equation}
\begin{eqnarray}\label{nonholoterm}
  &&\!\!\!{1\over M^3}\int {\rm d}^4\theta\,
 S S^\dagger \,
  (\Phi^2\Phi^\dagger+{\rm h.c.})\\
  &&\qquad\Rightarrow
  C_0\,(A^2A^*+{\rm h.c.})+
  {C_0\over F}\,AA^*\chi G+\dots\,.\nonumber
\end{eqnarray}
Because we are interested only in the power counting, we present these expansions schematically.  Each term implicitly includes a $\SUSY_V\times \SUSY_H$ breaking spurion analogous to $\alpha$.  The coefficients $m_0$, $m_{1/2}$, $A_0$ and $B_0$ are all of order $F/M$, which we take to be of order the weak scale $m_W$.  The coefficient $C_0$ is of order $F^2/M^3 \sim m_W^2/M$; that is, it is suppressed compared to the others by an additional power of $F/M^2\sim m_W/M$.  Note that the term
\begin{equation}
\int {\rm d}^2\theta\,
  S\,\Phi^2\Rightarrow  F\,A^2+
  A\,\chi G+\dots
\end{equation}
is not allowed by our power counting since it does not recover the larger $\SUSY_V \times \SUSY_H$ symmetry as $M \rightarrow \infty$.

\begin{figure}
\scalebox{0.65}
{\includegraphics{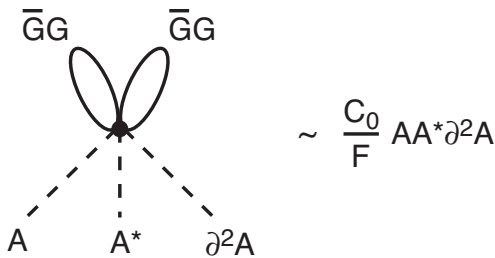}}
\caption{\label{fig1} New nonrenormalizable operators are generated by integrating out Goldstino loops.  The Goldstino loop momenta run to the scale $\sqrt F$.}
\end{figure}

The operators (\ref{softmass}) -- (\ref{nonholoterm}) are generically present in any theory of hidden sector supersymmetry breaking.  Their coefficients are determined by matching at the scale $M$.  In a given theory, however, one or more of the operators may be missing or suppressed because of physics at the scale $M$.

If a $\int {\rm d}^2\theta$ ``superpotential" term does not appear at the scale $M$, it will not be generated perturbatively by loops between $\sqrt F$ and $M$ because of the supersymmetric nonrenormalization theorem~\cite{Grisaru:1979wc}.  Therefore the corresponding operator will not appear in our effective theory.  By contrast, $\int {\rm d}^4\theta$ terms are not protected by the nonrenormalization theorem.  For that reason, we expect our effective theory to contain all $\int {\rm d}^4\theta$ operators consistent with the symmetries of the theory.  If one or more such operators are missing at the scale $M$, they will be generated by loops between $\sqrt F$ and $M$, with coefficients consistent with the power counting in $1/M$.

Note that in the effective theory, even in the absence of a superpotential, one can generate $\int {\rm d}^2\theta$ terms from supersymmetry breaking $\int {\rm d}^4\theta$ terms via the Giudice-Masiero mechanism~\cite{Giudice:1988yz}.  In particular, one can produce both supersymmetric,
\begin{equation}
  {1\over M}\int {\rm d}^4\theta\,
  S^\dagger \,\Phi^2
  \Rightarrow  {F \over M}\int {\rm d}^2\theta\,\Phi^2\,,
\end{equation}
and nonsupersymmetric operators,
\begin{equation}
  {1\over M^2}\int {\rm d}^4\theta\,
  S S^\dagger \,\Phi^2
  \Rightarrow  {F \over M^2}\int {\rm d}^2\theta\,
  S \,\Phi^2\,.
\end{equation}
Terms originating in this way are suppressed by an extra factor of $F/M^2$ relative to $\int {\rm d}^2\theta$ operators that arise directly at the scale $M$.

Below $\sqrt F$, the effective theory contains Goldstinos and visible sector fields.  In this effective theory, the rules for cutting off power divergent diagrams are the same as in our ETC toy model.  Goldstino loops are cut off at the scale $\sqrt F$, while loops involving visible sector fields run all the way up to the messenger scale $M$.  In fact, if the Goldstinos are completely integrated out, the effective theory of the visible sector fields is consistent to the scale $M$.  As in the ETC model, this is true even though the supersymmetry breaking condensate is generated at the lower scale $\sqrt F$.

\begin{figure}
\scalebox{0.65}
{\includegraphics{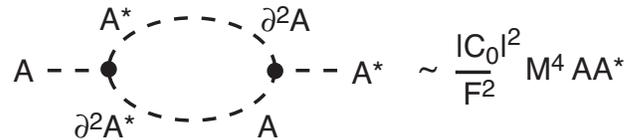}}
\caption{\label{fig2} Goldstino-induced operators renormalize the visible sector soft scalar masses.  The loop momentum runs to the scale $M$.}
\end{figure}

These rules for treating power divergent diagrams enforce the consistent power counting of the soft symmetry breaking terms.  For example, they imply that the suppression of the nonholomorphic trilinear term (\ref{nonholoterm}) relative to the others is required by the consistency of the low energy  theory.  The Goldberger-Treiman relation plays the critical role in the argument, because the expansion of the nonholomorphic term produces not only the trilinear coupling $C_0 A^2A^*$, but also a term with four Goldstinos,
\begin{equation}
  {C_0\over F^4}\, GG\overline G\overline G\,AA^* \partial^2 A\,.
\end{equation}
We now contract the Goldstino lines, as shown in Fig.~\ref{fig1}, and integrate the loops, each of which has a cubic divergence, up to $\sqrt F$.  This induces a new counterterm, which by consistency must
scale as
\begin{equation}\label{newcounterterm}
  {C_0\over F}\, AA^* \partial^2 A\,.
\end{equation}

Finally, we form the product of this term with its Hermitian conjugate and compute the one loop diagram shown in Fig.~\ref{fig2}, allowing the derivatives $\partial^2$ to act on the fields in the loop.  The cutoff of this quartically divergent diagram is $M$; the result is a correction to the scalar mass,
\begin{equation}\label{scalarcorr}
  {|C_0|^2 M^4\over F^2}\, A^*A\,.
\end{equation} 
If $C_0$ is of order $F^2/M^3$, as required by the power counting of the effective theory, the contribution to the soft mass is of order the weak scale,
\begin{equation}
  \delta m_A^2 \sim {F^2\over M^2}\sim m_W^2\,,
\end{equation}
which is not destabilized.
 
\section{Implications for the Minimal Supersymmetric Standard Model}

\begin{figure}
\scalebox{0.65}
{\includegraphics{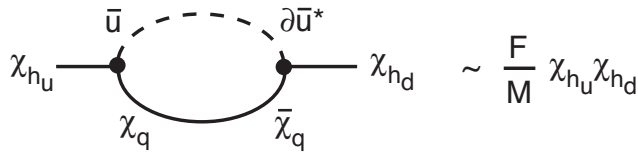}}
\caption{\label{fig3} The nonrenormalizable operator (\ref{muren}) gives rise to a supersymmetric $\mu$ term of order $F/M \sim m_W$.  The loop momentum runs to the scale $M$.}
\end{figure}

Let us now see what these considerations imply for the minimal supersymmetric standard model (MSSM).  The MSSM has the superpotential
\begin{equation}
P = \mu \,H_u H_d + \lambda_u\,Q\overline U H_u +
\lambda_d\,Q\overline D H_d\,.
\end{equation}
The soft supersymmetry breaking terms are generated at the matching scale $M$, and below $\sqrt F$ take the forms (\ref{softmass}) -- (\ref{nonholoterm}) and their variants at higher order.  In particular, these include the Higgs terms given by the operators
\begin{equation}
  {1\over M^2}\int {\rm d}^4\theta\,
  S S^\dagger \,(H_u^\dagger H_u + H_d^\dagger H_d)\,,
 \end{equation}
\begin{equation}
 {1\over M^2}\int {\rm d}^4\theta\,
  S S^\dagger \,H_u H_d\,,
\end{equation}
\begin{equation}\label{Aterm}
  {1\over M}\int {\rm d}^2\theta\,
  S\, (Q\overline U H_u +
Q\overline D H_d)\,,
\end{equation}
\begin{equation}\label{nonholo3}
{1\over M^3}\int {\rm d}^4\theta\,
 S S^\dagger \,
  (Q\overline U H_d^\dagger +
Q\overline D H_u^\dagger)\,.
\end{equation}
In the low energy effective lagrangian, each of these terms appears proportional to a spurion that breaks $\SUSY_V\times \SUSY_H\to \SUSY$.  Note that operator (\ref{Aterm}) produces the usual holomorphic $A$ term when one takes the $\theta\theta$ component of the expansion of $S$, while the operator (\ref{nonholo3}) is a higher dimension object that is suppressed and usually not included.

Let us use these terms to illustrate how ``missing" operators can be generated with the power counting as described above.  As before, we keep track only of powers of the scales $\sqrt F$ and $M$, not of loop factors of $4\pi$.  For our first example, we show how a weak scale $\mu$ term can arise even if it is missing at the messenger scale $M$.  To do this, we consider the Hermitian conjugate of the operator (\ref{nonholo3})  and extract the following term with four Goldstinos:
\begin{equation}
  {1\over F^4 M^3}\, GG\overline G\overline G\,\chi_{h_d} \sigma^m \bar{\chi}_{q} \partial_m \bar u^*\,.
\end{equation}
Contracting the Goldstino lines and integrating the Goldstino momenta to $\sqrt F$, we infer the existence of the counterterm
\begin{equation}\label{muren}
  { F \over M^3}\, \chi_{h_d} \sigma^m \bar{\chi}_{q} \partial_m \bar u^*\,.
\end{equation}
We now form the product of this operator with the ordinary Yukawa interaction, as shown in Fig.~\ref{fig3}.  Integrating the quark and squark momenta to $M$, we induce a supersymmetric Higgsino mass term of order $F/M \sim m_W$.  More specifically, the absence of a Higgsino mass of this size would require that this loop effect be canceled by a fine tuning at the messenger scale $M$.

\begin{figure}
\scalebox{0.65}
{\includegraphics{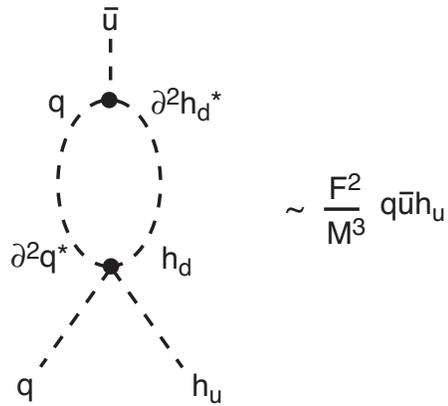}}
\caption{\label{fig4} Two nonrenormalizable operators generate a holomorphic scalar coupling suppressed by $F/M^2$.  The loop momentum runs to the scale $M$.}
\end{figure}

As a second example, we show that $\int {\rm d}^4\theta$ operators induce a trilinear $A$ term that is suppressed by $F/M^2$ with respect to the $\int {\rm d}^2\theta$ power counting.  For this we again use the higher order term (\ref{nonholo3}), which gives rise to the scalar operator
\begin{equation}
  {F \over M^3}\,  q \bar u \,\partial^2 h_d^*\,,
\end{equation}
exactly as in the discussion leading up to Eq.~(\ref{newcounterterm}).  In an analogous fashion, we also generate the operator
\begin{equation}
  {F \over M^4}\, q \partial^2 q^* h_u h_d
\end{equation}
from a higher order term in $1/M$,
\begin{equation}
   {1\over M^4}\int {\rm d}^4 \theta S S^\dagger Q Q^\dagger H_U H_D\,.
\end{equation}
Connecting them as shown in Fig.~\ref{fig4}, and integrating the squark momenta to $M$, we generate the holomorphic scalar trilinear $q \bar u h_u$, with coefficient of order $F^2/M^3 \sim m_W^2/M$.  Again, the precise statement is that a fine tuning at the messenger scale would be required to avoid such a term in the low energy lagrangian.  Of course, this term is suppressed by $m_W/M$ compared to the weak scale; the result is interesting, even in principle, only in the absence of the lower order $\int {\rm d}^2\theta$ operator (\ref{Aterm}) (which does {\it not\/} requiring a fine tuning).

Finally, we emphasize that the correct power counting in $1/M$ is essential for the consistency of the low energy theory.  In particular, if one operator is arbitrarily assigned an anomalously large coefficient, Goldstino loop effects will drive up the other coefficients to restore the proper counting.

As a phenomenologically important example, consider the operator (\ref{nonholo3}), which gives rise to nonholomorphic trilinear terms such as $q \bar u h_d^*$.  Hall and Randall~\cite{Hall:1990ac} and Weinberg~\cite{Weinberg:2000cr} pointed out that such terms, of the form $m_W A^2A^*$, are ``soft'' in the sense that they do not destabilize the hiercharchy in a theory (such as the MSSM) that has no gauge singlets.  This observation is correct, but incomplete, since the Goldberger-Treiman relation introduces Goldstino couplings that {\it do\/} destabilize the hierarchy in the presence of such a term.

To see that this is so, suppose that the operator (\ref{nonholo3}) is included with an arbitrary coefficient $C_0$.  By an argument identical to the one that leads to Eq.~(\ref{scalarcorr}), there must then be a counterterm of the form
\begin{equation}\label{nonholoMSSM}
  {|C_0|^2 M^4\over F^2}\, (h_u^* h_u + h_d^* h_d)  \,.
\end{equation} 
Because the MSSM contains no gauge singlets, the authors of Refs.~\cite{Hall:1990ac} and~\cite{Weinberg:2000cr} propose to take $C_0$ to be of order $F/M \sim m_W$.  However, we see from Eq.~(\ref{nonholoMSSM}) that if $C_0\sim m_W$, the Higgs mass correction from Goldstino loops is of order
\begin{equation}
  \delta m_{h_u,h_d}^2 \sim M^2\,.
\end{equation}
In other words, Goldstino loops destabilize the weak scale if one does not use the correct power counting for the symmetry breaking terms.  Consistency requires that $C_0$ must be taken to be of order $F^2/M^3$, implying that nonholomorphic trilinear terms must have coefficients of order $m_W^2/M$, rather than $m_W$.

We close by emphasizing the very general nature of this result.  Just as the interactions with the Goldstone bosons follow from the form of chiral symmetry breaking in extended technicolor, the interactions with Goldstinos follow immediately, via the Goldberger-Treiman relation, from the form of the soft terms in the MSSM.  The two are inextricably linked because they share the physics of symmetry breaking and its communication through the messenger sector.  The special power of effective field theory, when combined with nonlinear realizations, is that this connection is revealed in a way that is simultaneously transparent, elegant and useful for phenomenology.

%\vspace{.25truein}

\acknowledgments

This paper is dedicated to the memory of Julius Wess, from whom, both personally and through his published work, we learned so much.  We are grateful to the Aspen Center for Physics, where portions of this work were completed.  This work was supported in part by the National Science Foundation under grant NSF-PHY-0401513.

\end{document}